\pdfoutput=1

\documentclass[12pt]{iopart}
\usepackage{iopams}
\usepackage{color,graphicx,rotating, cite, hyperref}

\begin{document}

\title[Parallelizing quantum circuit synthesis]{Parallelizing quantum circuit synthesis}

\author{Olivia Di Matteo$^{1,2}$ and Michele Mosca$^{2,3,4,5}$}

\address{$^1$ Department of Physics and Astronomy, University of Waterloo, Waterloo, Canada}
\address{$^2$ Institute for Quantum Computing, Waterloo, Canada}
\address{$^3$ Department of Combinatorics and Optimization, University of Waterloo, Waterloo, Canada}
\address{$^4$ Perimeter Institute for Theoretical Physics, Waterloo, Canada}
\address{$^5$ Canadian Institute for Advanced Research, Toronto, Canada}

\ead{\mailto{odimatte@uwaterloo.ca}, \mailto{mmosca@uwaterloo.ca}}
\vspace{10pt}

\begin{abstract}
Quantum circuit synthesis is the process in which an arbitrary unitary operation is decomposed into a sequence of gates from a universal set, typically one which a quantum computer can implement both efficiently and fault-tolerantly. As physical implementations of quantum computers improve, the need is growing for tools which can effectively synthesize components of the circuits and algorithms they will run.  Existing algorithms for exact, multi-qubit circuit synthesis scale exponentially in the number of qubits and circuit depth, leaving synthesis intractable for circuits on more than a handful of qubits. Even modest improvements in circuit synthesis procedures may lead to significant advances, pushing forward the boundaries of not only the size of solvable circuit synthesis problems, but also in what can be realized physically as a result of having more efficient circuits.

We present a method for quantum circuit synthesis using deterministic walks. Also termed pseudorandom walks, these are walks in which once a starting point is chosen, its path is completely determined. We apply our method  to construct a parallel framework for circuit synthesis, and implement one such version performing optimal $T$-count synthesis over the Clifford+$T$ gate set. We use our software to present examples where parallelization offers a significant speedup on the runtime, as well as directly confirm that the 4-qubit 1-bit full adder has optimal $T$-count 7 and $T$-depth 3.

\end{abstract}

%
\vspace{2pc}

\section{Introduction} \label{sec:intro}

Quantum computers, like their classical counterparts, will require a compiler which can translate from a human-readable input or programming language into operations which can be executed directly on quantum hardware. Circuit synthesis is an integral part of the compilation process. Given an arbitrary quantum circuit $C$ and a universal gate set 
$\mathcal{G}$, one seeks to find a decomposition
\begin{equation}
 U_k U_{k-1} \cdots U_2 U_1 = C, \quad U_i \in \mathcal{G},
 \label{eq:synthesis}
\end{equation}
\noindent where $k$ represents the depth of the circuit. A myriad of algorithms currently exist to find such a decomposition \cite{Dawson2006, KMM2013, Selinger2015, KMM2016, Bocharov2013, Ross2015, Bocharov2015, Wiebe2014, Welch2016, KMM2013Exact, Giles2013, MITM, Gosset2014, Kliuchnikov2015, Forest2015}. They are generally divided into two classes, those which synthesize approximately (i.e. $|| U_k \cdots U_1 - C || < \epsilon$) and others which synthesize exactly. Some procedures work for a single qubit, whereas others have been generalized to multiple qubits. Most of these algorithms were designed to work over the Clifford+$T$ universal gate set, though other gate sets such as the $V$-basis have also been studied ~\cite{Ross2015, Bocharov2013}.

Many of the algorithms which perform exact synthesis fall victim to the fact that the time and space  used depend exponentially on both the number of qubits and the depth of the circuit in question. Even on a reasonably fast machine, synthesis of circuits with more than a handful of qubits and layers of depth becomes intractable. 

In this work, we propose a method of circuit synthesis based on a heuristic search technique commonly used in cryptanalysis: collision finding based on deterministic, or pseudorandom walks. These are walks through a search space such that once a starting point is chosen, the path is completely determined. More generally, we show how we can use deterministic walks to traverse the space of possible circuits of a given depth and find solutions to the synthesis problem. A key ingredient in our method is a mapping from the unitary operators constructed from the gate set $\mathcal{G}$ to binary strings of a constant length, and a suitable mapping back to the set of unitary operators. When such mappings are defined, we can synthesize circuits over any universal gate set, on any number of qubits, by applying any existing walk method which can search the space. 

The structure of this article is as follows. We begin in Section \ref{sec:walks} with a discussion of deterministic walks, and how we can map quantum circuit synthesis to these types of problems. The subsequent sections pertain to our choice of implementation of one such method, namely parallel circuit synthesis. In Section \ref{sec:framework} we briefly lay out the procedure for parallel synthesis and provide a runtime complexity estimate, detailing the important parameters which affect the scaling of our algorithm. Section \ref{sec:implementation} introduces our software implementation, pQCS, which performs optimal $T$-count synthesis using parallel search. Section \ref{sec:results} contains the numerical results of large-scale experiments run on a Blue Gene/Q supercomputer. Here we showcase the significant advantages afforded to us by parallelization. We conclude in Section \ref{sec:conclusion} and suggest avenues of future research on this topic.

\section{Walking through circuits}

\label{sec:walks}

Consider a hash function $h : \mathcal{D} \rightarrow \mathcal{R}$, typically considered to operate over binary strings. If $h$ is a good hash function, then for an arbitrary input $x \in \mathcal{D}$, the value $h(x) = y \in \mathcal{R}$ will be in practice indistinguishable from a random output. Suppose there exists another function $r : \mathcal{R} \rightarrow \mathcal{D}$, unrelated to $h$, which maps elements of its range back to the domain (such a function is commonly termed a \emph{reduction function}). Repeatedly applying $r \circ h$ to an input will produce a trail of points scattered throughout $\mathcal{D}$. However, once the initial input is chosen, the progression of the trail is completely determined, hence we favour the term \emph{deterministic} rather than pseudorandom walk even though the path of the walk appears random due to the natures of $h$ and $r$. 

Such determinism has led to a set of algorithms with a variety of applications. One well-known variation is rainbow tables ~\cite{Oeschlin2003}, which are used for finding pre-images of hash functions (conventionally with the intention of cracking passwords). Collision finding in one hash function, or claw finding between two functions has also been accomplished in parallel using deterministic walks ~\cite{vOW1}, and was used to find collisions in double DES ~\cite{vOW2}.

Deterministic walks are advantageous due to their low storage requirement: one need only store the starting point of a walk, its ending point, and the number of intermediate steps, whereas conventional search techniques would store the value of every point computed throughout.

\begin{figure} [htbp]
 \centering
 \includegraphics[scale=0.55]{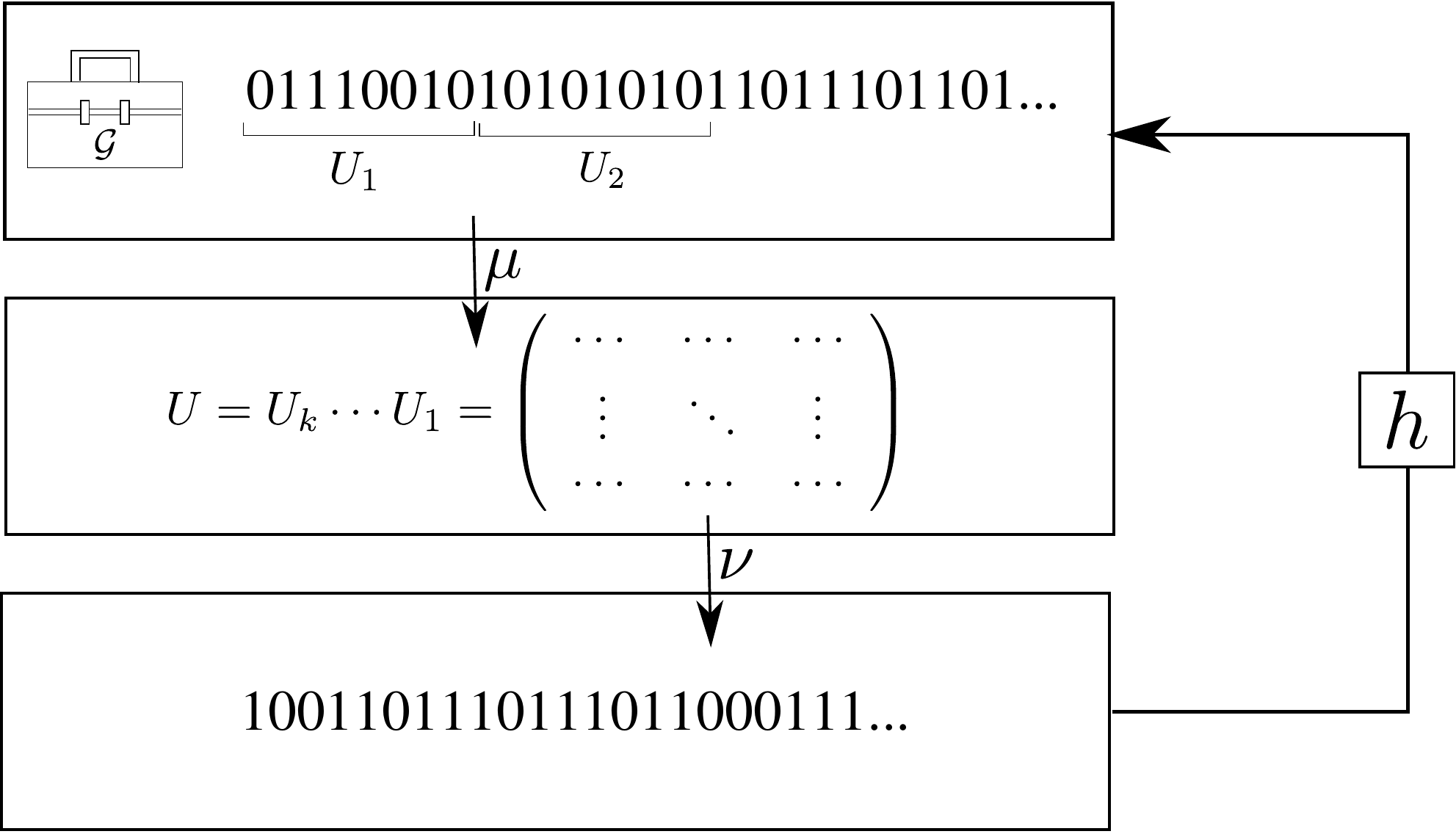} 
 \caption{A schematic diagram showing the process of walking over circuits. Binary strings are mapped to products of unitary matrices over the gate set $\mathcal{G}$ via some correspondence $\mu$. The product of these matrices is then mapped via $\nu$ back to a binary string, which is then passed through a hash function $h$. Repeated application of $h \circ \nu \circ \mu$ allows us to traverse the set of possible circuits in a pseudorandom fashion.}
 \label{fig:circuit_walk_schematic}
\end{figure}

With this in mind, we show how one can map the problem of circuit synthesis to a problem that can be solved using an algorithm based on deterministic walks. We have, as per (\ref{eq:synthesis}), a product constructed from the universal gate set $\mathcal{G}$. It is possible to specify a unique way of encoding the information about $\{U_1, \ldots, U_k\}$ into binary strings $\{\mathbf{b}_1, \ldots, \mathbf{b}_k\}$ of equal length $\ell$ (where we assume $\ell$ is sufficiently long as to encompass all the information described in what follows). Suppose $\mathcal{G}$ contains a number of single- and two-qubit gates.  If we enumerate all the gates in $\mathcal{G}$, then for each $U_i$ we might use a few bits to identify all the constituent gates, and maybe a few more to specify if we should use their Hermitian conjugates. We will also need to indicate on which qubit(s) they act. Furthermore, there must be some space to indicate controls and targets where appropriate. Given any gate set, we can find a way of doing this such that every possible $U_i$ can be represented by a unique string $\mathbf{b}_i$. Then, the concatenation $(\mathbf{b}_k|\cdots |\mathbf{b}_1)$ will be a unique string of length $k \ell$ representing the product of unitaries $U_k \cdots U_1$.

We can perform a deterministic walk over unitary matrices as follows; this process is displayed graphically in Figure \ref{fig:circuit_walk_schematic}. Let us define a function $\mu$ which maps a binary string of length $k \ell$ to a unitary matrix over a specified gate set $\mathcal{G}$. Then define a mapping $\nu$ from the unitaries over $\mathcal{G}$ back to binary strings $\{0, 1 \}^*$. Finally, choose a good hash function $h$ from $\{0, 1\}^*$ to strings of length $k \ell$ (this may be a simple hash function, or a combination of hash and reduction-type functions). Repeatedly applying $h \circ \nu \circ \mu$ to a randomly chosen binary string of length $k \ell$ will allow us to traverse products of unitaries in a pseudorandom fashion; we can then use this to search the space of possible solutions to  (\ref{eq:synthesis}).

\section{Parallel circuit synthesis} 

\label{sec:framework}

Once we have mappings as proposed in Section \ref{sec:walks}, we can reformulate the circuit synthesis problem as a problem which can be solved using search algorithms based on deterministic walks. We specifically implemented one which performs parallel claw finding. Let $h_1: \mathcal{D}_1 \rightarrow \mathcal{R}$ and $h_2 : \mathcal{D}_2 \rightarrow \mathcal{R}$ be two hash functions. A \emph{claw} between $h_1$ and $h_2$ is a pair of inputs $x_1 \in \mathcal{D}_1, x_2 \in \mathcal{D}_2$ such that
\begin{equation}
 h_1(x_1) = h_2(x_2).
\end{equation}
\noindent This is, in a sense, a collision search between two functions.

Our interest in claw finding stems from recent work on circuit synthesis using a meet-in-the-middle (MITM) approach ~\cite{MITM}. The motivation for that work is as follows. One can of course find a decomposition of (\ref{eq:synthesis}) by brute force, computing all possible combinations starting from depth 1 up until a solution is found. Let $\xi$ represent the number of unitaries having depth 1. Typically $\xi$ will depend exponentially on the number of qubits, $n$. Then, the runtime for brute force synthesis of a circuit with depth $k$ takes time $\mathcal{O}(\xi^{k})$. A MITM approach achieves a roughly square-root speedup over this, accomplished by dividing the synthesis equation in half:
\begin{equation}
  U_{\lceil \frac{k}{2} \rceil} \cdots U_1 = U^\dag_{\lceil \frac{k}{2} \rceil + 1 } \cdots U^\dag_k C, \quad U_i \in \mathcal{G}. \label{eq:mitm}
\end{equation}
\noindent Databases of unitaries having the form of each side of  (\ref{eq:mitm}) are sequentially constructed (starting from depth 1), stored in binary trees, and then searched through until a suitable decomposition is found. This reduces the size of the search space by a square root factor, yielding runtime  $\mathcal{O} \left(\xi^{\lceil\frac{k}{2} \rceil} \log \left( \xi^{\lceil\frac{k}{2} \rceil} \right) \right)$, where the log factor is picked up due to the binary search.

To parallelize circuit synthesis, we build on the principles of the MITM algorithm. Rather than searching through static binary trees, we search the space in parallel, adapting a search technique originally developed for cryptanalysis ~\cite{vOW1}. Though our runtime will retain the exponential dependence on $n$ and $\lceil k / 2 \rceil$, it scales inversely with the number of processors, allowing us to tackle larger problems which were infeasible using previous methods, as well as speed up the synthesis of some known circuits. We provide a brief description of the algorithm here as it pertains specifically to circuit synthesis. For a more detailed description, the reader is referred to ~\cite{vOW1} or ~\cite{DiMatteo2015}.

Recall (\ref{eq:mitm}), and for simplicity, let us define
\begin{eqnarray}
 V &:=& U_{\lceil \frac{k}{2} \rceil} \cdots U_1,\\
  W &:=& U^\dag_{\lceil \frac{k}{2} \rceil + 1 } \cdots U^\dag_k C,
\end{eqnarray}
\noindent as representing the left and right sides of this equation. Define a suitable mapping between unitary matrices and binary strings of length $k \ell$ as in Section \ref{sec:walks}. Then let $\mathcal{V}^\prime$ represent the set of binary strings that are of the form $V$, and likewise  $\mathcal{W}$ those of the form $W$. When $k$ is odd, $\mathcal{V}^\prime$ and $\mathcal{W}$ may differ in size by a factor of $\xi$. In this case, we partition $\mathcal{V}^\prime$ into equal sized chunks $\mathcal{V}^\prime_0, \ldots, \mathcal{V}^\prime_{\xi - 1}$, and consider $\mathcal{V} = \mathcal{V}^\prime_i$ independently (a search can then be executed with each $\mathcal{V}^\prime_i$ sequentially or in parallel, adding another layer of parallelism to the implementation).  When $k$ is even, we simply let $\mathcal{V} = \mathcal{V}^\prime$.  

Let $\mathcal{N} = \{0, 1 \}^{k \ell}$. Define functions $z_1 : \mathcal{N} \rightarrow \mathcal{N}$ and $z_2:\mathcal{N} \rightarrow \mathcal{N}$. One way these functions might be implemented is by converting the input string into a sequence of unitary matrices (in $\mathcal{V}$ for $z_1$ and $\mathcal{W}$ for $z_2$), computing their product, deriving a new binary string with the information about each of the matrix elements, and then running that string through a known hash function so that the outputs of both functions are in the same space and in practice appear to be random. 

Let us define a `super' function $f : \mathcal{N} \times \{1, 2\} \rightarrow \mathcal{N} \times \{1, 2\}$ such that one application of $f$ is a single step in the deterministic walk, i.e. $f(x, b) = z_b(x)$. Finding a claw between $z_1$ and $z_2$ is now equivalent to finding a collision in $f$ with distinct values for $b$, i.e. we must find two inputs $x_1$ and $x_2$ such that
\begin{equation}
 f(x_1, 1) = f(x_2, 2).
\end{equation}

Consider $m$ processors all having access to a shared memory.  We will denote some fraction $\theta$ of points in $\mathcal{N}$ as marked, or distinguished. Every processor chooses a random starting pair $(n_0, b_0)$ in $\mathcal{N} \times \{1, 2\}$. Repeatedly applying $f$ produces a \emph{trail} through the space of possible circuits, which roughly half the time  will produce a part of (\ref{eq:mitm}) which is an element of $\mathcal{V}$, and the other half of the time will produce an element of $\mathcal{W}$.  The trail continues until the next input, say $x_d$, is a distinguished point. The trail is then terminated.

The collection of found distinguished points is stored in the shared memory. Distinguished points are stored as a triple consisting of the first pair $(n_0, b_0)$, the last pair $(n_d, b_d)$, and the value $d$, which is the number of steps taken to reach the distinguished point. When a processor finishes its trail, it will attempt to add its distinguished point to the shared memory. If it sees that a trail ending at the given point is not present in this shared memory, it will insert it and then begin a new trail. However, if it sees that there is already a triple in storage which ended at the same distinguished point but had a \emph{different} starting point, it means that somewhere along the way these two trails must have merged. The processor then takes the starting points of these two trails, and traces back through them to locate the merge point.

\begin{figure}[h]
 \centering
 \includegraphics[scale=0.3]{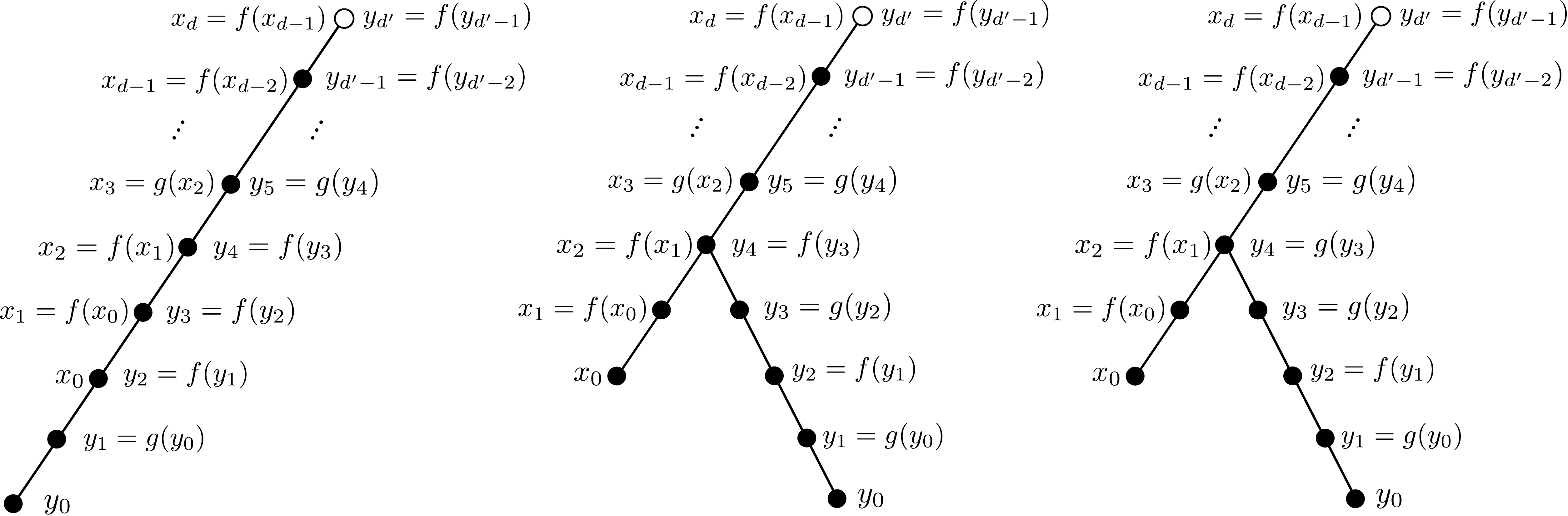}
 \caption{Possible ways two trails can merge. Let $f$ and $g$ be two functions between which we want to find a claw. (Left) One trail starts before the other. (Centre) The two trails merge after performing the same function, i.e. a collision $f(x_1) = f(y_3)$. (Right) The two trails merge after performing a different function, i.e. a claw $f(x_1) = g(y_3)$.}
 \label{fig:trail_schematic}
\end{figure}

There are a number of possibilities here, as depicted in Figure \ref{fig:trail_schematic}. First, it could be that one trail started ``before'' the other, i.e. the merge point was at the beginning of the shorter trail. Another possibility is that when the trails merged, both had just performed $z_1$, or both had just performed $z_2$. Even if the inputs were different, this case does not provide us with a solution to the problem at hand. The final case is that immediately before they merged, one trail performed $z_1$ and one performed $z_2$; it is only in this final case that we have found a solution. With the information about the inputs in the step just before the collision, we can extract the unitary matrices from the binary string, and have fully synthesized our circuit.

The runtime complexity of this algorithm can be estimated by applying the parameters of our problem directly to that in ~\cite{vOW1}. The size of the spaces $\mathcal{V}^\prime$ and $\mathcal{W}$ are 
\begin{equation}
 N_{\mathcal{V^\prime}} = \xi^{\left\lceil \frac{k}{2} \right\rceil}, \quad
 N_{\mathcal{W}} = \xi^{\left\lfloor \frac{k}{2} \right\rfloor}. 
\end{equation}
Our algorithm then scales as
\begin{equation}
 T_{QCS} \propto \xi^{\left\lceil \frac{k}{2} \right\rceil + \frac{1}{2} \left\lfloor \frac{k}{2} \right\rfloor} \frac{1}{\sqrt{w}} \frac{1}{m} \tau, \label{eq:runtime}
\end{equation}
\noindent where $w$ is the number of distinguished points that can be held in memory. The parameter $\tau$ is the execution time for a single iteration of $z_1$ or $z_2$, the bulk of which will likely be spent performing matrix multiplication. Let us assume in the worst case that we are taking the product of $\lceil \frac{k}{2} \rceil$ $2^n \times 2^n$ unitaries using a multiplication algorithm which scales as $(2^n)^{\alpha }$, where $\alpha$ is some constant, typically $2 < \alpha \leq 3$. Thus, we obtain our final estimate
\begin{equation}
  T_{QCS} \propto 2^{\alpha n} \xi^{\left( \left\lceil \frac{k}{2} \right\rceil + \frac{1}{2} \left\lfloor \frac{k}{2} \right\rfloor \right) } \frac{1}{\sqrt{w}} \frac{1}{m} \left\lceil \frac{k}{2} \right\rceil. \label{eq:runtime_withmatrixmult}
\end{equation}

\noindent As previously mentioned, this time is still exponential in the number of qubits as well as the depth of the circuit. We also note that it is often the case that matrix multiplication can be parallelized, or that some specific properties of the implementation at hand (such as sparsity) can be leveraged so as to improve the scaling. What is key here is that the runtime benefits from being inversely proportional to the number of processors and available memory.

\section{Implementation details} \label{sec:implementation}

\subsection{Optimal $T$-count synthesis}

The synthesis algorithm we chose to apply our approach to is the optimal $T$-count algorithm presented in ~\cite{Gosset2014}. Such an algorithm is relevant as in many state-of-the-art  methods for fault-tolerant quantum computation, $T$ gates are considered to be expensive to implement due to the need to distill magic states (see, for example, \cite{Fowler2012}). 

Let $\mathcal{P}_n$ represent the $n$-qubit Pauli group. We reshuffle and rewrite the decomposition of a circuit $C$ as 
\begin{equation}
 e^{i \phi} R(P_t) \cdots R(P_1) D = C,
 \label{eq:tcount_synthesis}
\end{equation}
where $t$ is the $T$-count, $D$ is a Clifford, $P_j \in \mathcal{P}_n$, and 
\begin{equation}
 R(P_j) = \frac{1}{2} \left( 1 + e^{\frac{i \pi}{4}} \right) I_{2^n} +  \frac{1}{2} \left( 1 - e^{\frac{i \pi}{4}} \right) P_j.
\end{equation}
It thus suffices to find a set of $t$ Paulis and a Clifford which will satisfy (\ref{eq:tcount_synthesis}) up to a global phase. The dependence on the global phase can also be removed by using the channel representation of every matrix in the above equation:
\begin{equation}
 \widehat{R(P_t)} \cdots \widehat{R(P_1)} \widehat{D} = \widehat{C},
\end{equation}
\noindent where the channel representation of some matrix $U$ is the matrix with coefficients 
\begin{equation}
 \widehat{U}_{ij} = \frac{1}{2^n} \hbox{Tr} \left( P_i U P_j U^\dag \right), \quad P_i, P_j \in \mathcal{P}_n.
\end{equation}

\noindent The channel representation of an $n$-qubit unitary has dimension $4^n \times 4^n$, with each row and column being indexed by a Pauli operator.

Using the optimal $T$-count algorithm has afforded us with a number of advantages. First of all, the $T$-count formulation  allows us to represent each unitary matrix in the sequence as a list of $n$-qubit Paulis. With binary symplectic representation we can then represent each Pauli directly as a binary string, which leads to a very simple mapping with which we can perform our deterministic walks. Another strong point of the algorithm is that the channel representations of $R(P)$ for $P \in \mathcal{P}_n$ are sparse matrices. Thus, we were able to implement a sparse matrix multiplication algorithm which allows us to very quickly compute most matrix products, despite the channel representations having dimension $4^n \times 4^n$.

We can apply (\ref{eq:runtime}) and (\ref{eq:runtime_withmatrixmult}) to the optimal $T$-count synthesis to obtain a runtime estimate. Each $R(P)$ contributes a single $T$ gate to the circuit, and can be considered as a single layer of depth in this implementation. Thus, we have that $\xi = 4^n -1$, as all Paulis save for the identity are valid choices. Our estimate for the runtime is thus
\begin{equation}
 T_{QCS-T} \propto 2^{n \left(2 \alpha + 2  \left\lceil \frac{t}{2} \right\rceil +  \left\lfloor \frac{t}{2} \right\rfloor \right) } \frac{1}{\sqrt{w}} \frac{1}{m} \left\lceil \frac{t}{2} \right\rceil.
\end{equation}

\subsection{Computer specifications}

\begin{sidewaysfigure}
 \vspace{16cm}
 \includegraphics[scale=0.5]{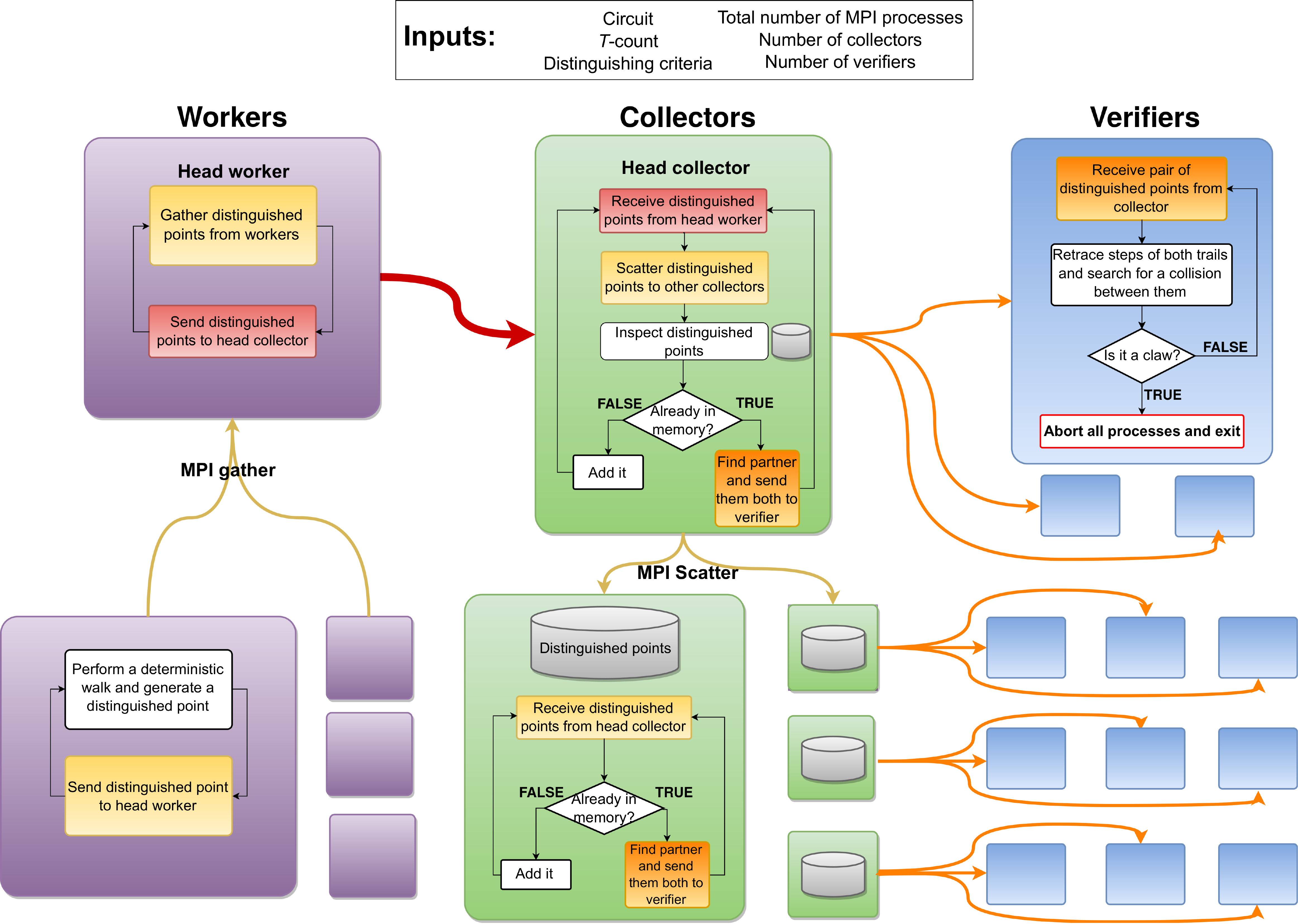} 
 \caption{A flowchart of work distribution for the version of pQCS run on the Blue Gene/Q. Worker processors perform walks and generate distinguished points. These are funneled through a head worker  to a head collector processor, which then distributes the points amongst all the collectors for processing and storage. Each collector has access to a number of verifiers. Collectors which find pairs of walks ending at the same distinguished point distribute the pairs to their verifiers to check for a claw.}
 \label{fig:flowchart}
\end{sidewaysfigure}

  We implemented the optimal $T$-count version of the parallel algorithm in C++11. It is called pQCS (\textbf{p}arallel \textbf{q}uantum \textbf{c}ircuit \textbf{s}ynthesis), and is available for download and research use at \href{https://qsoft.iqc.uwaterloo.ca/\#software}{https://qsoft.iqc.uwaterloo.ca/\#software}. Parallelization was accomplished using the Boost.MPI compiled library \cite{Boost}. A scaled down version of pQCS which uses only OpenMP for parallelization (and can be run on a standard multi-core personal computer) is also available in the above package.

pQCS was extensively tested on two large-scale machines. The OpenMP-only version was tested on SHARCNET's Orca using a single node with up to 16 processors at 2.2GHz speed. The MPI version was tested on Scinet's Blue Gene/Q (BG/Q) supercomputer, which has 65536 processors at 1.6GHz speed. The largest test we have run to date involved a total of 8192 cores. All results below are from trials on the BG/Q. A flowchart and description of the distribution of work in the MPI version is presented in Figure \ref{fig:flowchart}.

\section{Results} \label{sec:results}

\subsection{Determining effective simulation parameters} \label{subsec:parameters}

pQCS has a number of tunable parameters. In what follows we will synthesize a known circuit, the Toffoli gate, and explore the scaling of our algorithm.

In the original description of the parallel collision finding algorithm \cite{vOW1}, each processor was responsible for performing not only the search for a distinguished point, but also storing it and subsequently checking the validity of any possible solutions; it is from this setup that the heuristic runtimes are derived. In pQCS, however, processors are divided into three categories (as per Figure \ref{fig:flowchart}) which communicate via MPI. Worker processors perform deterministic walks and generate distinguished points. Distinguished points are collected and stored in-core on collector processes. Each collector has access to a number of verifier processors, to which pairs of walks are sent for verification when the possibility of a claw occurs. The parameters $m$ and $w$ may not necessarily depend then on the total number of processors, but rather on the number of processors in one or more of the different classes. For example, $w$ will depend solely on the number of collectors, whereas we expect $m$ to be a function of the number of workers, assuming a sufficient number of collectors and verifiers are in place.

\begin{figure} [htbp]
 \includegraphics[scale=0.55]{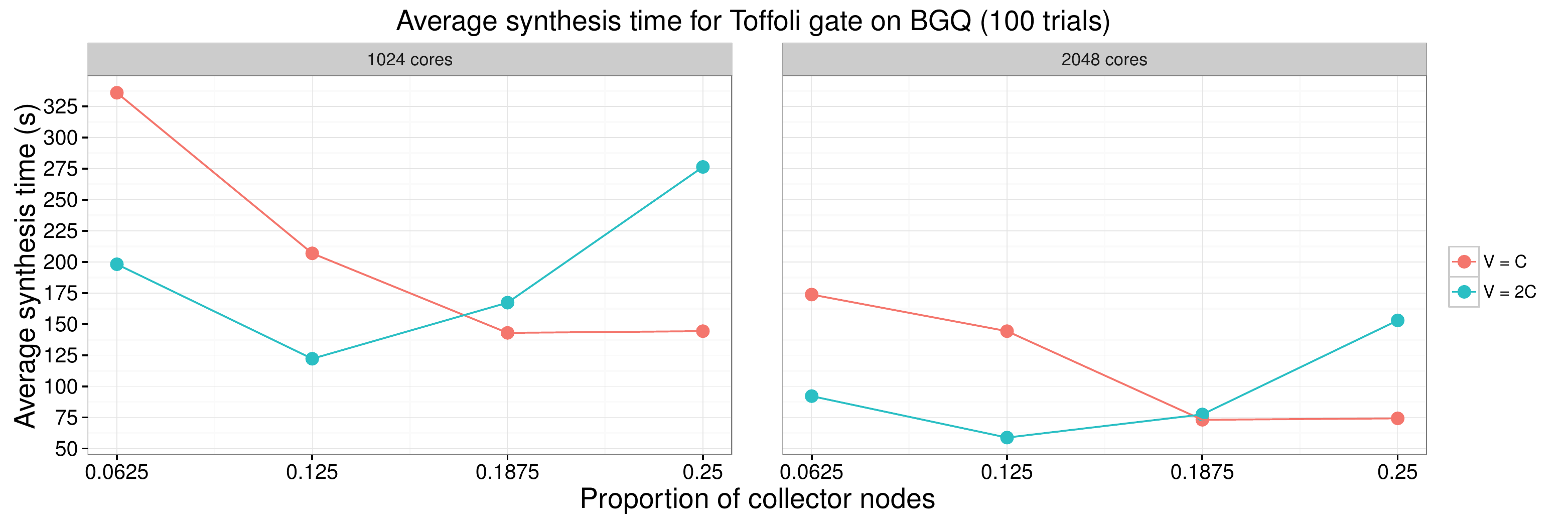} 
 \caption{Variation of the number of collectors and verifiers when synthesizing the Toffoli gate. The fraction of distinguished points was set at 1/4. The legend $V = C$ indicates equal amounts of collectors and verifiers, whereas $V = 2C$ indicates two verifiers per collector. We find that the optimal number of collectors seems to be about 1/8 the total number of processors, and the number of verifiers to be twice that, at 1/4 the total number.}
 \label{fig:tof_collector_count}
\end{figure}

First, we focus on how many collectors and verifiers we should use. We chose two values for the total number of cores, 1024 and 2048.  We then varied the fraction of nodes designated as collectors in increments of 1/16, from 1/16 to 1/4 the total (values outside this range clearly yielded inferior results). For each fraction of collectors, we either used the same, or double the number of verifiers. The results of these trial runs are shown in Figure \ref{fig:tof_collector_count}. In all these trials we let 1/4 of the points in the space be designated as distinguished (later we will fine-tune this parameter as well). Each point is the average of 100 independent trials. We find that for both total quantities of processors, the optimal number of collectors is 1/8 the total number, and for verifiers 1/4 the total. When more than 3/8 of the total processes are being used on storage and verification, there are not enough workers to perform the deterministic walks. On the other hand, when there are too many workers, each collector must store and process a larger collection of distinguished points each time. Furthermore, more time will be spent by the workers gathering and sending the increased quantity of distinguished points.

\begin{figure} [htbp]
 \centering
 \includegraphics[scale=0.5]{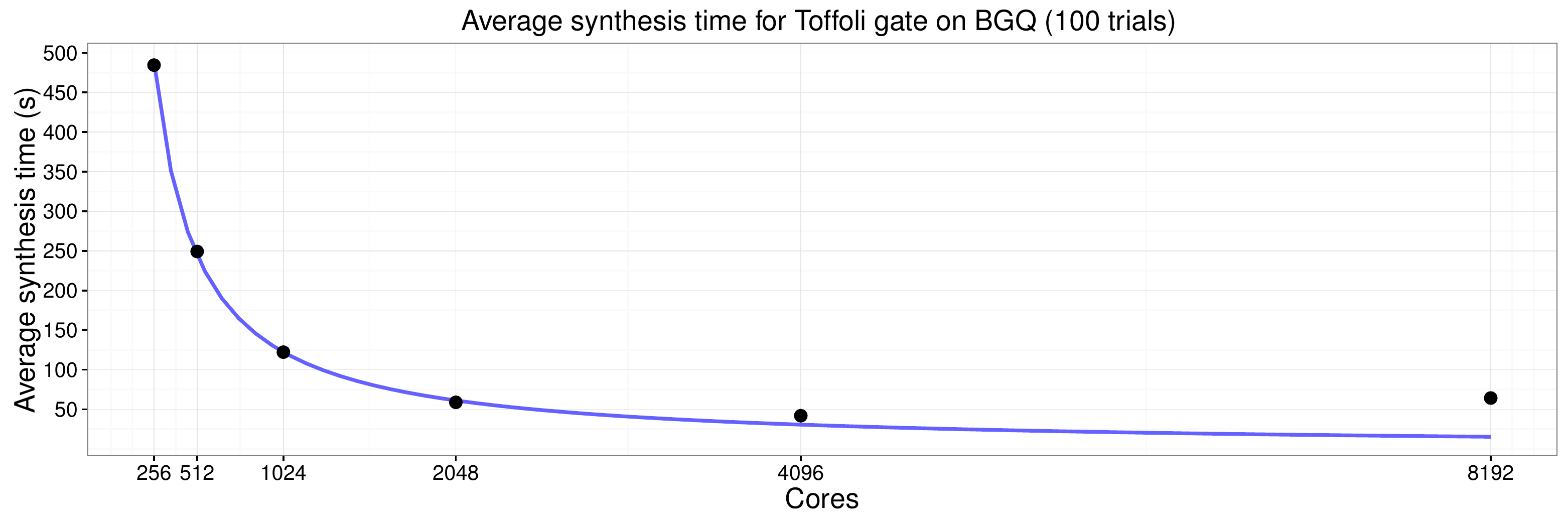} 
 \caption{Varying the total number of cores when synthesizing the Toffoli gate. We used 512 collectors and 1024 verifiers with 1/4 of the points distinguished. Data follows the inverse trend line quite closely until around the 4096 core mark. After this point, it is likely that the overhead and communication costs are too large for a problem of this size.}
 \label{fig:tof_cores}
\end{figure}

With this knowledge, we then tested the Toffoli with varying number of cores. Again, we let 1/4 of the points be distinguished and take the average of 100 independent trials. The results are shown in Figure \ref{fig:tof_cores}. We see clearly here the expected inverse dependence on the number of processors as predicted by  (\ref{eq:runtime}). We do note that there is significant deviation from the expected trend when we reach 8192 cores. We suspect that for a problem of this size, the parallel overhead and communication costs outweigh the potential benefits of using this many cores.

\begin{figure} [htbp]
 \centering
  \includegraphics[scale=0.5]{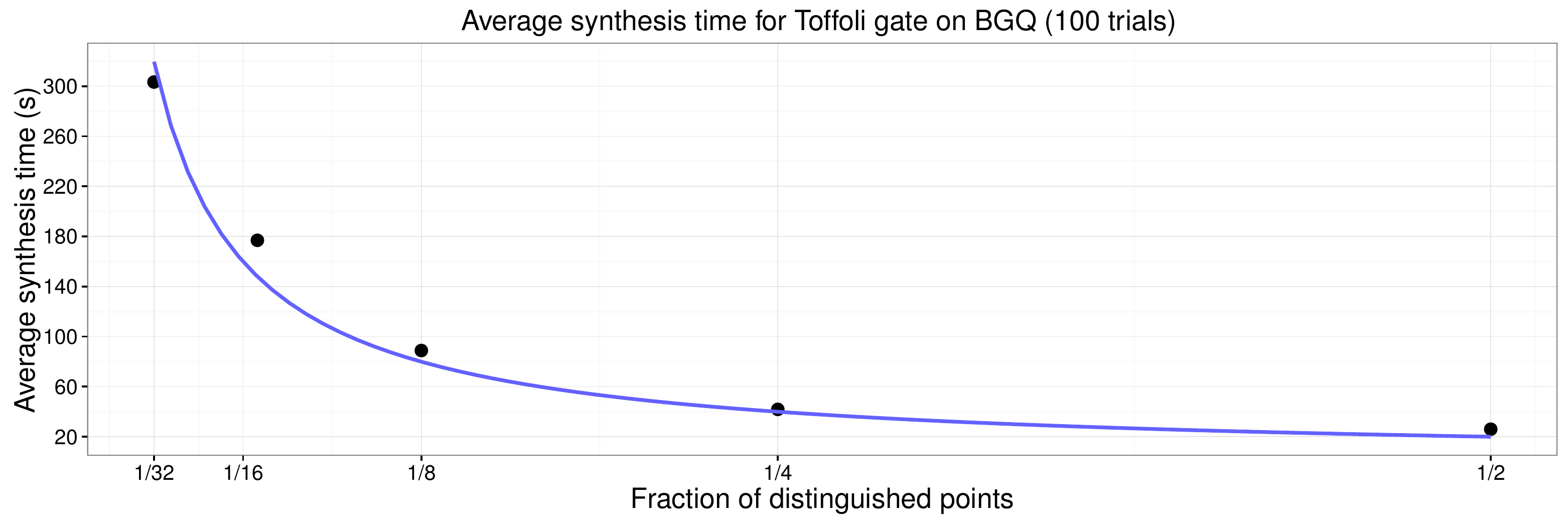} 
  \caption{Varying the fraction of distinguished points while synthesizing the Toffoli gate. As the size of the search space is much less than the available memory, we see roughly the expected inverse dependence on the fraction of distinguished points.}
  \label{fig:tof_dp_fraction}
\end{figure}

 Finally, we investigate how the runtime varies with the fraction of distinguished points, $\theta$. In the case of the Toffoli, the amount of available memory using the above number of processors on the BG/Q is significantly greater than that required to store even the entire space. Variation of this parameter is thus somewhat contrived for such a (relatively) small problem. In this case we would expect an inverse dependence on $\theta$ (see the Appendix for more details). We ran 100 trials on 4096 processors (512 collectors and 1024 verifiers) using fractions of distinguished points $\{$1/2, \enskip 1/4, \enskip  1/8, \enskip 1/16,\enskip  1/32$\}$.  The results are displayed in Figure \ref{fig:tof_dp_fraction}, where we see the expected inverse dependence.  We also report here our best synthesis times for the Toffoli gate, clocking in at roughly 26s on average. To fully explore the effects of this parameter (and more importantly the dependence on the available memory $w$), we would need to use a much larger circuit.

\subsection{Benchmarking known circuits}

\begin{figure} [htbp]
\centering
\includegraphics[scale=0.7]{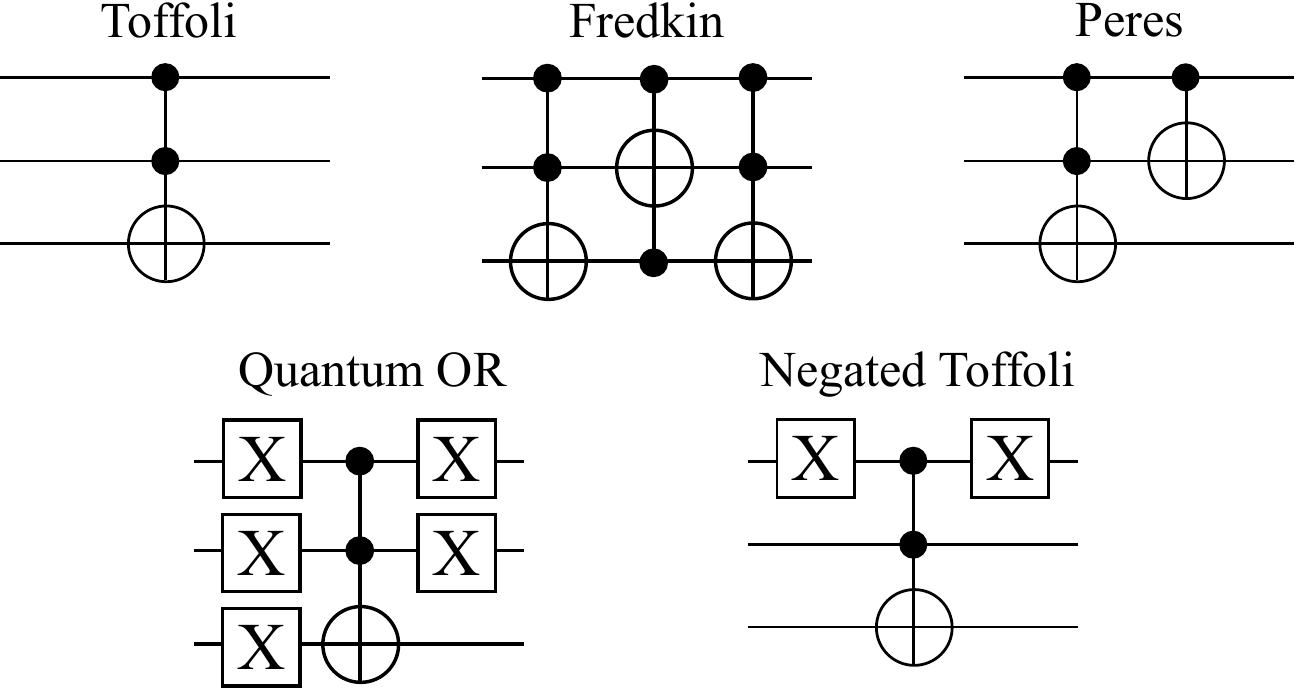} 
\caption{Circuit diagrams for the five 3-qubit circuits with $T$-count 7 which we synthesized.}
\label{fig:tc7_collection}
\end{figure}

Some of the largest circuits which were directly synthesizable by both the original MITM algorithm and optimal $T$-count algorithm were those with $T$-count 7 on $3$ qubits ~\cite{MITM, Gosset2014}. There are a number of such circuits, shown in Figure \ref{fig:tc7_collection}. Using our knowledge from optimization of parameters in the previous section (4096 cores, 1/2 points distinguished, 512 collectors and 1024 verifiers), we obtain the synthesis times reported in Table \ref{tab:3qtc7}. We note that at roughly 25s, these times are a marked improvement over those reported in \cite{DiMatteo2015}, which were greater than 4 minutes. This highlights the advantage of using many processors, and is a promising sign that we will be able to synthesize circuits which are much larger in a reasonable amount of time.

\begin{table}[h]
  \centering

  \begin{tabular}{|c|c|c|}
  \hline
  \textbf{Circuit} & \textbf{Average time (s)} & \textbf{Std. dev.  (s)} \\ \hline \hline
  Toffoli & 25.9870 & 11.0733 \\ \hline
  Fredkin & 25.0031 & 9.4869 \\ \hline 
  Peres & 25.4931 & 11.1753 \\ \hline
  Quantum OR & 24.1854 & 9.1417 \\ \hline
  Negated Toffoli & 26.9162 & 11.1561 \\ \hline
  \end{tabular}
  \caption{Synthesis of a known set of 3-qubit circuits all having optimal $T$-count 7. All results come from 100 independent trials using 4096 cores (512 collectors, 1024 verifiers), and 1/2 of points distinguished as per the results of Section \ref{subsec:parameters}.}
        \label{tab:3qtc7}
\end{table}

\subsection{Pushing the boundaries}

\begin{figure} [htbp]
\centering
\includegraphics[]{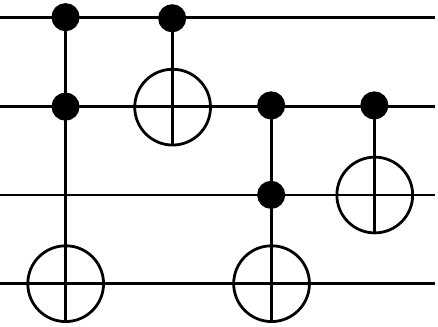} 
\caption{The 4-qubit adder. We find directly that it has $T$-count 7 and $T$-depth 3, and that these results are optimal.}
\label{fig:adder} 
\end{figure}

The largest circuit synthesized to date using pQCS  is the 4-qubit 1-bit full adder, shown in Figure \ref{fig:adder}. A synthesized version of this adder appeared in ~\cite{MITM} with $T$-count 8, where it was accomplished using peephole optimization techniques. It was suspected that it has $T$-count 7 \cite{MattConvo}, which we confirm.

\begin{figure} [htbp]
 \centering
 \includegraphics[scale=0.27]{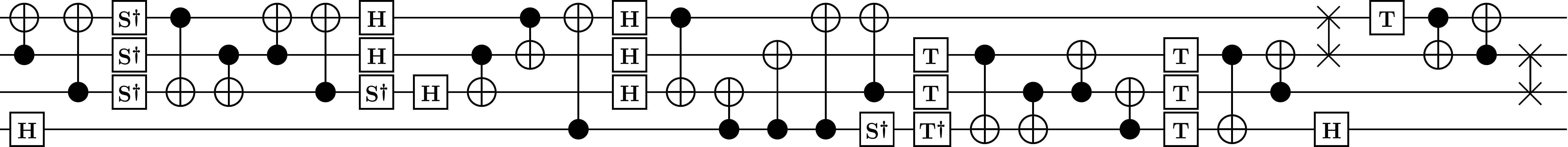} 
 \caption{A decomposition of the 4-qubit adder over Clifford+$T$, optimized for $T$-depth. The X gates indicate swaps.}
 \label{fig:adder_full}
\end{figure}

 The first successful synthesis of the adder took 12.5 hours using 4096 cores (512 collectors, 1024 verifiers) and 1/2 points distinguished. We note that a circuit as large as the adder would likely benefit from a larger number of processors, and so more testing is in progress. A full version of the circuit is shown in Figure \ref{fig:adder_full}. The initial output of pQCS is a sequence of Paulis and a unitary corresponding to a Clifford gate as per (\ref{eq:tcount_synthesis}). The Pauli portion of the circuit ($R(P_7)\cdots R(P_1)$) was generated using the algorithm given in the appendix of ~\cite{Gosset2014}, and the Clifford component was generated using the algorithm in ~\cite{Aaronson2004}. The resultant sequence of gates was then optimized for $T$-depth using $T$-par ~\cite{Tpar}. Interestingly, this new synthesis of the adder led to the observation that it requires identical resources as the Toffoli gate, i.e. $T$-count 7, $T$-depth 3, and to the question of whether this is a coincidence. In fact, it was subsequently pointed out to us that this adder is affine equivalent to the Toffoli (i.e. unitarily equivalent up to application of CNOTs) ~\cite{TomConvo}.

\section{Concluding remarks} \label{sec:conclusion}

We have presented a framework for quantum circuit synthesis based on deterministic walks, as well as an algorithm and software for parallel quantum circuit synthesis. We have observed a clear advantage over existing techniques using a relatively modest number of processors, and were able to directly synthesize a 4-qubit circuit which would have been intractable using previous methods. 

Ongoing and future work on pQCS includes improvements to the application structure and parallelization routines, extensions for synthesis in general over a specified gate set, and the implementation of approximate circuit synthesis. Furthermore, we seek to push the application to its limits in order to fully characterize the scaling, in particular with respect to the available memory once the circuit search spaces become sufficiently large.

\ack
We thank SHARCNET and Scinet for use of their computing resources. Computations were performed on the SOSCIP Consortium'€™s Blue Gene/Q computing platform. We are grateful to Barbara Collignon from IBM for support with BG/Q development. We also thank Matt Amy, Vadym Kliuchnikov, and Tom Draper for helpful discussions. Funding was provided by NSERC, CIFAR, and SOSCIP. IQC and the Perimeter Institute are supported in part by the Government of Canada and the Province of Ontario. SOSCIP is funded by the Federal Economic Development Agency of Southern Ontario, the Province of Ontario, IBM Canada Ltd., Ontario Centres of Excellence, Mitacs and 15 Ontario academic member institutions. A portion of this work was completed while attending the Quantum Computer Science workshop at the Banff International Research Station (17-22 April 2016).

\appendix

\section*{Appendix}
\setcounter{section}{1}
The runtimes presented in (\ref{eq:runtime}) and (\ref{eq:runtime_withmatrixmult}) stem from a so-called `flawed' runtime analysis originally presented in \cite{vOW1}. Suppose we are searching for a collision in a space of size $N$, and that the available memory is full with $w$ distinguished points. The number of steps required to find a single collision in this case is

\begin{equation}
 \frac{N \theta}{w} + \frac{2}{\theta}, 
 \label{app_eq:flawed_runtime}
\end{equation}
where $\theta$ is the fraction of points which are distinguished. The first term comes from the fact that to fill the memory with $w$ distinguished points, $w / \theta$ elements in the space will be traversed on average, and any given point in a new trail has a $1/N$ probability of landing on a previously seen point; the second term comes from the need to trace back through both trails to locate the collision, and each trail has length $1/\theta$ on average. 

The assumption is made that there is a single `golden' collision. In this case $N/2$ `bad' collisions will be found on average before the golden one is found. If we parallelize using $m$ processors and assume each step in a trail takes time $\tau$, then we obtain a runtime
\begin{equation}
 T \propto \frac{1}{m} \left( \frac{N^2 \theta}{2 w} + \frac{ N}{\theta} \right) \tau
\end{equation}

The next step taken in \cite{vOW1} is to differentiate and find $\theta$ such that (\ref{app_eq:flawed_runtime}) is optimized, which is what results in the inverse-square-root dependence on $w$. They then performed computational experiments for a range of $w$ and $N$ in order to find optimal prefactors.

However, the optimal $\theta$ is expressed in terms of $w / N$, which when $w >> N$ (as is the case when we synthesize the Toffoli on the BG/Q) would not result in a fractional $\theta$. So let us continue a hypothetical analysis of this form without finding the optimal $\theta$. Consider the case where we are optimizing for $T$-count. In the most general case, the two halves of the MITM equation will be different sizes $N_1$ and $N_2$ where $N_1 = 4^{n \lceil \frac{t}{2} \rceil }$ and $N_2 = 4^{n \lfloor \frac{t}{2} \rfloor}$ ($t$ being the $T$-count). Since when we have an odd depth we partition the larger space and search sequentially (in theory this could also be done in parallel), we must add a prefactor of $N_1 / N_2$ in front of the runtime, and the $N$ becomes $2 N_2$, because the full space we're searching is that of $N_2 \times \{1, 2\}$. As for $\tau$, let's assume $\tau = \lceil \frac{t}{2} \rceil 4^{\alpha n}$ where $\alpha$ is a constant which reflects the complexity of the matrix multiplication algorithm. Then we have that

\begin{eqnarray}
 T & \propto & \frac{1}{m} 4^{n \left( \lceil \frac{t}{2} \rceil - \lfloor \frac{t}{2} \rfloor \right)} \left( \frac{4^{2n \lfloor \frac{t}{2} \rfloor + 1} \theta}{2w} + \frac{2 \cdot  4^{n \lfloor \frac{t}{2} \rfloor}}{\theta} \right)    \left\lceil \frac{t}{2} \right\rceil 4^{\alpha n}\\
  &=& \left\lceil \frac{t}{2} \right\rceil  \frac{1}{2 m} 4^{n \left( \alpha + 1 + \lceil \frac{t}{2} \rceil \right)} \left( \frac{4^{n \lfloor \frac{t}{2} \rfloor} \theta}{w} + \frac{1}{\theta} \right) 
\end{eqnarray}

When $w >> N_2$, the first term disappears and the expression reduces to
\begin{equation}
 T \propto \left\lceil \frac{t}{2} \right\rceil  \frac{4^{n \left( \alpha + 1 + \lceil \frac{t}{2} \rceil \right)}}{m \theta},    \\
\end{equation}

\noindent which is exponential in both $n$ and $t$, and inversely proportional to both $m$ and $\theta$, precisely what we have observed in practice. 

\end{document}